\documentclass[aps,onecolumn,prd,showpacs,showkeys,preprintnumbers,superscriptaddress,nobibnotes,notitlepage,floatfix,longbibliography,nofootinbib]{revtex4-2}
\usepackage[normalem]{ulem}
\usepackage{graphicx}
\usepackage{amsmath}
\usepackage{amssymb}
\usepackage{xcolor}
\usepackage[colorlinks=true,citecolor=blue,urlcolor=blue]{hyperref}

\def\be{\begin{equation}}
\def\ee{\end{equation}}
\def\bea{\begin{eqnarray}}
\def\eea{\end{eqnarray}}

\def\mev{\, {\rm MeV}}
\def\kev{\, {\rm keV}}
\def\ev{\, {\rm eV}}
\def\g{\, {\rm g}}

\newcommand{\nhz}{\rm nHz}

\newcommand{\gsim}{\lower.7ex\hbox{$\;\stackrel{\textstyle>}{\sim}\;$}}
\newcommand{\lsim}{\lower.7ex\hbox{$\;\stackrel{\textstyle<}{\sim}\;$}}

\begin{document}

\title{Primordial black holes from Q-balls produced in a first-order phase transition}

\author{James B.~Dent}
\affiliation{Department of Physics, Sam Houston State University, Huntsville, TX 77341, USA}

\author{Bhaskar Dutta}
\affiliation{Mitchell Institute for Fundamental Physics and Astronomy, Department of Physics  and Astronomy, Texas A$\&$M University, College Station, Texas 77843,  USA}

\author{Jason Kumar}
\affiliation{Department of Physics and Astronomy, University of Hawai'i, Honolulu, HI 96822, USA}

\author{Danny Marfatia}
\affiliation{Department of Physics and Astronomy, University of Hawai'i, Honolulu, HI 96822, USA}

\begin{abstract}
We consider the formation of Q-balls in false vacuum remnants during a cosmological first-order
phase transition.  We find that under certain circumstances Q-balls can collapse to form primordial black holes.  This scenario can produce multimessenger signals that may be observed at upcoming 
experiments, including $1\text{--}100~\nhz$ gravitational waves from the phase transition, and gamma-rays emitted from primordial black holes as Hawking radiation and as superradiance.  These signals are quite distinctive, and differ markedly from signals expected from Fermi-balls.
The reheating of the dark sector from the phase transition may address the Hubble tension.
\end{abstract}

\maketitle

\section{Introduction}

Cosmological phase transitions in the early Universe can produce a variety of phenomena which can be probed with observations, including gravitational waves, baryogenesis, and dark matter production.  
The false vacuum itself can leave remnants in the form of topological defects, such as domain walls 
and cosmic strings.
In a recently studied scenario, a different type of false vacuum remnant can survive well after the phase transition. A long-lived particle is essentially confined to the false vacuum because its mass in the true vacuum is much larger than in the false vacuum.  As these particles collect in shrinking bubbles of false vacuum, they may enter a condensed state and even collapse into a primordial black hole (PBH).  
This scenario has been studied in the case where the particle confined to the false vacuum is a fermion
and the non-topological soliton is a 
Fermi-ball~\cite{Hong:2020est,Kawana:2021tde,Marfatia:2021hcp}. 
In this work, we consider the case in which the particle is a boson. 

When a long-lived particle is confined to the false vacuum, its density grows as the bubble shrinks, 
providing a new method of generating a large overdensity as a seed for PBH formation.  As the density grows quantum effects become important when the occupancy of low-lying states becomes ${\cal O}(1)$.
There are then two possibilities for the behavior of the overdensity at low temperatures, depending on whether the particle is a long-lived boson or a fermion~\cite{Kawana:2022lba}. 

If long-lived scalar particles are confined to the false vacuum, they can condense into a Bose-Einstein condensate (BEC), creating a Q-ball.  
Q-balls provide unique and interesting phenomenology because they have a much weaker pressure than Fermi-balls to counteract the negative pressure arising from the vacuum energy.  
This yields a very different size for the Q-ball, and in turn a very different mass distribution for PBHs from their collapse.\footnote{In Ref.~\cite{Cotner:2019ykd}, Q-balls were considered as PBH progenitors arising from a fragmentation process under certain conditions on the scalar potential. The scalar field in that scenario acquires a large vacuum expectation value at the end of inflation and the scalar lumps formed from the fragmentation may cluster with a sufficient overdensity to form a PBH. The formation mechanism and associated cosmology differ from that considered in the present work.}  In particular, since the Q-ball can be much smaller than a Fermi-ball before it collapses, 
it can produce a much smaller PBH mass density 
while yielding a 
similar gravitational wave signal from the first-order phase transition (FOPT).

We find that the most interesting region of parameter space is for FOPTs that occur at a temperature in the $100~\kev - 1~\mev$ range.  In this range, the FOPT can produce gravitational waves in the $1-100~\nhz$ range which can be observed 
with upcoming observatories, such as THEIA.  Moreover, the PBHs produced by the collapse of the Q-ball can survive past recombination, and the Hawking radiation produced as they evaporate 
may be detected at AMEGO-X. Thus, this scenario provides a multi-messenger signal which is clearly distinguishable from 
the case of a fermion condensate.  Morever, the reheating of the dark sector after the phase transition can provide 
a contribution to $N_{\rm eff}$ which may alleviate the Hubble tension.

This work is situated  within a larger class of studies on PBH production from trapped remnants (see also \cite{Hong:2020est,Kawana:2021tde,Marfatia:2021hcp,Kawana:2022lba}).\footnote{Other PBH formation mechanisms include, but are not limited to,  delayed~\cite{Liu:2021svg,Kawana:2022olo} and supercooled transitions~\cite{Lewicki:2023ioy,Gouttenoire:2023naa}, or bubble collisions~\cite{Jung:2021mku} and inverted bubble collapse~\cite{Murai:2025hse}.} However, as we have mentioned and will explore in detail below, a scalar condensate produces a parameterically smaller PBH mass than its fermionic counterparts, giving a unique combination of observational possibilities.

The plan of this paper is as follows.  In Section~\ref{sec:condensate}, we describe the process by which 
scalar particles can be trapped in bubbles of false vacuum, condense into a Q-ball, and finally collapse to a PBH.  
In Section~\ref{sec:quartic}, we consider a specific example in which the FOPT arises from 
a quartic thermal effective potential.  In Section~\ref{sec:constraints}, we describe multimessenger signals 
of this scenario.  In Section~\ref{sec:results}, we present our results, including the parameter space that can be probed with future observations.   We discuss these results in Section~\ref{sec:discussion}, 
and conclude in Section~\ref{sec:conclusion}.

\section{Formation of a PBH from  a Q-ball}  
\label{sec:condensate}

First, we briefly review the process by which particles (denoted $X$) collect in false vacuum bubbles during a dark FOPT.  We consider a scenario in which the mass $m_X$ depends on the 
vacuum expectation value of a scalar field, which differs between the high and low temperature 
phases. Specifically, $m_X$ is smaller in the false vacuum (i.e., the high temperature phase), than in the true vacuum (i.e., the low temperature phase).  
Then, as bubbles of the true vacuum nucleate, the $X$ particles get trapped in the false vacuum.  Eventually, 
the true vacuum dominates, leaving disconnected shrinking false vacuum bubbles that contain all the $X$ particles.  At the time of percolation, the false vacuum bubbles are disconnected, so that $X$ cannot move from one false vacuum bubble to another.

To describe the formation of false vacuum bubbles, we follow the discussion of Refs.~\cite{Lu:2022paj,Marfatia:2024cac}.
We define the percolation temperature $T_*$ as the temperature at which the fraction of the Universe left in the false vacuum is $F(T_*) = 1/e$.   If we assume a monochromatic radius distribution for the false vacuum bubbles, then 
the false vacuum bubble radius at percolation is given by
\bea
R_{f*} &=& \left(\frac{3 v_w}{4\pi \Gamma_*} \right)^{1/4}\,,
\eea
where $v_w$ is the speed of the bubble wall and $\Gamma_*$ is the true vacuum bubble nucleation rate per volume at 
$T_*$.  
However, the radius distribution of false vacuum bubbles at percolation will not be monochromatic and can be expressed 
in terms of $R_{f*}$ as 
\bea
\frac{dn}{dR_f}(t_*) &=& \frac{1}{32 R_{f*}^4} x^4 e^{-x} (1 - e^{-x})\,,
\label{eq:dndr}
\eea
where $n$ is the number density of false vacuum bubbles and 
$x = \exp \left[6^{1/4} R_f / R_{f*} \right]$.  The average false vacuum bubble radius is 
$\langle R_f \rangle \sim  0.84 R_{f*}$ and the total number density of false vacuum bubbles 
is $n \sim 0.11 R_{f*}^{-3}$.  The distribution of bubble radii is flat for small $R_f$, but falls sharply for $R_f \gtrsim R_{f*}$.

If the particles $X$ are their own antiparticles, then they generally can simply annihilate away in the false vacuum bubbles.  But if they are not self-conjugate, a primordial asymmetry can exist between particles and antiparticles which we parameterize by $\eta_X$. Mechanisms that produce $\eta_X$ are discussed in the appendix of Ref.~\cite{Hong:2020est}.
If we let $n_X - n_{X^*} = \eta_X s(T)$ be the net number density of $X$, where $s(T)$ is the entropy density in the false vacuum, then the net number of $X$ in a false vacuum bubble of radius $R_f$ 
is given by
\bea
Q &=& \frac{\eta_X s(T_*)}{F(T_*)} \frac{4\pi}{3} R_f^3\,.
\label{eq:QofR}
\eea

\subsection{Q-ball formation}

Once the fraction of the volume in the false vacuum is sufficiently small, and the false vacuum 
bubbles are disconnected, the net number of long-lived $X$ particles in each bubble is fixed.    
But the bubble will continue to shrink until the pressure differential between the true and false vacuum 
vanishes.  
To determine the radius $R$ at which the false vacuum bubble is stabilized, we should solve the equation 
$\partial \Delta F / \partial V =0$, where $\Delta F$ is the difference in Helmholtz free energy between the 
true vacuum and false vacuum bubble.  This equation implies 
that the pressure differential across the bubble wall is zero.

We denote by $U_0 (T) (>0)$ the difference between the thermal effective potential in the false vacuum (equivalent to the false vacuum free energy density) and in the true vacuum at temperature $T$.
Initially, the $X$ particles will form a classical relativistic ideal gas at temperature $T$, with a 
free energy given by $F_X \sim -3QT \log [RT/Q^{1/3}] -QT$.  We then find
\bea
\frac{\partial \Delta F}{\partial R} &=& -\frac{3QT}{R} + 4\pi U_0 (T) R^2\,, 
\eea
yielding a bubble radius $R = (3QT / 4\pi U_0(T))^{1/3}$.  
Since we find that $Q$ is typically $\mathcal{O}(10^{60})$, we are justified in 
neglecting the tension of the bubble wall.  We also neglect any contribution to the potential energy 
arising from self-interactions of the bosons.
At this stage, it does not matter if the 
$X$ are bosons or fermions, only that the number in each bubble is fixed.

As $T$ decreases and the bubble shrinks, the density of $X$ particles will increase.  When the 
expected occupancy of the low-lying states becomes ${\cal O}(1)$, the $X$ will instead form a quantum gas.
At this point, we can contrast the cases of a bosonic and fermionic particle.  For the case of a fermion, 
each of the low-lying states will be occupied by one particle up to the Fermi energy ($\propto Q^{1/3}/R$), yielding a total kinetic energy
$\propto Q^{4/3} / R$.

For the case in which $X$ is a boson, when the occupancy of the ground state is ${\cal O}(1)$, 
the behavior of the false vacuum bubble must be solved numerically.
But once the temperature drops below the Einstein condensation temperature, 
most of the particles will condense into the ground state, and the behavior of the false vacuum bubble can again be determined  analytically.  This object is a Q-ball.

For relativistic bosons, the density of states scales as ${\cal D}(\epsilon) \propto \epsilon^2$, where $\epsilon$ is the energy of the state.  At large $\epsilon$, for which the density of states is insensitive to the boundary conditions, we find ${\cal D}(\epsilon)/V = (1/2\pi^2) \epsilon^2$, 
where $V$ is the volume.  
At temperature $T$, the occupancy of the state is then 
given by $f(\epsilon, \mu, T) = 1/\left(\exp[(\epsilon -\mu)/T]-1 \right)$, where $\mu$ is the chemical 
potential which ensures that the total number of particles is $Q$.

We consider the limit in which the temperature has cooled down sufficiently, so that the occupancy of the ground state is $\gg 1$.  For spherical bubbles, the energy of the ground state is $\epsilon_0 = \pi /R$.
If the occupancy of the ground state is large, then $(\epsilon_0 - \mu)/T \ll 1$. 

The number of particles in excited states is 
\bea
N_{\rm exc} &=& \int_{>\epsilon_0}^\infty d\epsilon ~  {\cal D}(\epsilon) f(\epsilon, \mu, T) 
= \frac{2}{3\pi} R^3 \int_{>\epsilon_0}^\infty d\epsilon \frac{(\epsilon - \epsilon_0)^2}{\exp[(\epsilon - \epsilon_0)/T ] -1}\,,
\nonumber\\
&=& \frac{4 \zeta(3)}{3\pi} R^3 T^3\,.
\eea
We may then define the Einstein condensation temperature, $T_{\rm EC}$, as the temperature at which $N_{\rm exc} = Q$, 
yielding 
\bea
T_{\rm EC} &=& \left(\frac{4 \zeta(3)}{3\pi} \right)^{-1/3} \frac{Q^{1/3}}{R} \,. 
%\propto Q^{1/12} [U_0 (T)]^{1/4}\,.
\eea
Below this temperature, the occupancy of the ground state becomes large.

We may then write the kinetic energy in thermal equilibrium, as 
\bea
E_{\rm kin} (T) &=& Q \epsilon_0 + \int_{>\epsilon_0}^\infty d\epsilon ~ (\epsilon -\epsilon_0) {\cal D}(\epsilon) f(\epsilon, \mu, T)\,,
\eea
where the integral accounts for particles in excited states.  Taking the density of states to be 
${\cal D} (\epsilon) = C R^3 (\epsilon -\epsilon_0)^2$, where $C = 2/3\pi$ for the case of a spherical bubble of radius $R$, we 
find at small $T$,
\bea
E_{\rm kin} (T) &\sim& Q \epsilon_0 + \frac{2}{3\pi} R^3 \int_{>\epsilon_0}^\infty d\epsilon ~ 
\frac{(\epsilon - \epsilon_0)^3}{\exp[(\epsilon - \epsilon_0)/T]-1 } 
= \frac{\pi}{R }Q  + \frac{2\pi^3}{45} R^3 T^4\,.
\eea

Most scalar particles are in the condensed phase, and their entropy is negligible.
However, the small number of scalar particles in the normal 
phase (i.e., in excited states) can still contribute to the entropy and pressure.
We  have 
\bea
F &=& \frac{\pi}{R }Q
%\frac{c_0}{R }Q  
- \frac{2\pi^3}{135} R^3 T^4 + \frac{4\pi}{3} R^3 U_0 (T)\,,
\eea
where the first two terms yield a positive pressure due to the kinetic energy of the scalars, and the 
third term yields a negative pressure due to the false vacuum bubble energy density.

Note that, for $T \gtrsim U_0^{1/4} (T)$, there is no equilibrium solution for $R$.  This indicates that 
the pressure due to particles in the normal phase alone is sufficient to resist the negative pressure from 
the vacuum energy.  If the scalars were indeed in a scalar condensate, then the pressure from the normal phase component 
would cause the bubble to expand, decreasing $T_{\rm EC}$ until $T > T_{\rm EC}$.  This 
indicates that occupancy of the ground state will not become large until $(\pi^2/90) T^4 \lesssim U_0 (T)$.

Once the false vacuum bubble cools to $T \ll [U_0(T)]^{1/4}$, we have 
\bea
R_{\rm Q-ball} &=& Q^{1/4}  \left[4 U_0 (T) \right]^{-1/4}\,,
\nonumber\\
M_{\rm Q-ball} &=&  \frac{4}{3} \pi Q^{3/4} \left[4  U_0 (T) \right]^{1/4} = 
1.07 \times 10^{19}\g
\left[\frac{\left[ U_0 (T) \right]^{1/4}}{\mev} \left(\frac{Q}{10^{60}} \right)^{3/4} \right]\,. 
\label{eq:MofQ}
\eea
Denoting the mass of $X$ in the true vacuum by $M_{X,\rm true}$, we may choose 
\bea
\frac{\partial M_\mathrm{Q-ball}}{\partial Q} &=& \pi Q^{-1/4} \left[4 U_0 (T) \right]^{1/4} < M_{X, \rm true}\,. 
\eea
Since we also have 
\bea
\frac{\partial^2 M_\mathrm{Q-ball}}{\partial Q^2} &=& 
-\frac{3\pi}{4}  Q^{-5/4} \left[  U_0 (T) \right]^{1/4} < 0 \,,
\eea
the Q-ball is indeed stable against decay and fragmentation.

Note that in the case where the particles confined to the false vacuum are fermions, we instead 
find $M_{\rm Fermi-ball} \propto Q [U_0(T)]^{1/4}$~\cite{Kawana:2021tde}.  In that case, the energy scales linearly with the number of 
particles because Pauli exclusion implies that the kinetic energy grows rapidly with the number of particles, 
increasing the pressure.  
For the case of bosons, the energy of the false vacuum bubble grows much more slowly with the number of 
particles, since almost all the bosons will lie in the ground state, at low temperatures.  
In particular, 
we see that a Q-ball will be much lighter than a Fermi-ball.

\subsection{The coupling between $X$ and $\phi$}

The coupling between $X$ and $\phi$ in the false vacuum leads to a weak Yukawa interaction  
which will cause the $X$ condensate to eventually collapse into a PBH.  
But this coupling must also lead to a large difference in the mass of $X$ between the 
true and false vacuum, which is responsible for the $X$ particles being confined to the 
false vacuum.  However, it is important to note that the coupling between $X$ and $\phi$ can look 
very different when expanded about the true vacuum than when expanded about the false vacuum.  
To illustrate this point, we represent an example of the coupling between $X$ and $\phi$ as 
\bea
{\cal L} &=& \ldots - X^* X w(\phi) \,,
\label{eq:X_phi}
\eea
where $w(\phi)$ is a mass dimension-2 function of $\phi$.  We adopt a notation in which 
the false vacuum is at $\langle \phi \rangle = 0$ and the true vacuum is at 
$\langle \phi \rangle = v$.  We take $\phi$ to have the power series form, 
\bea
w(\phi) &=& \Lambda^2 \sum_{n=0}^\infty a_n ({\phi /\Lambda})^n \,,
\eea
where the  $a_n$ are dimensionless with $|a_n| < 1$, and $\Lambda$ is the energy scale below 
which the effective Lagrangian provides a valid description of interactions between 
$X$ and $\phi$ in the false vacuum; we only need $\Lambda \gtrsim 1/R$.  It is important to note that the  
scales $\Lambda$ and $v$ are unrelated to each other because $v$ is the vev of the $\phi$ in the true vacuum, and can be chosen so that $v \gg \Lambda \gtrsim 1/R$.  There is 
no particular reason why the interaction in Eq.~(\ref{eq:X_phi}) should provide a perturbative description 
of interactions between the $X$ and $\phi$ at the energy scale $v$.

In the false vacuum, we may take $m_X^2 \sim a_0 \Lambda^2$ to be very small.  Similarly, 
we need the effective Yukawa coupling $g \sim a_1 \Lambda (R / m_X)^{1/2}$ to be quite small, which can be achieved 
by taking $a_1$ sufficiently small.
Furthermore, we ensure that the presence of the condensate with 
$\langle X^* X \rangle \sim (U_0 Q)^{1/2}$ in the false vacuum does not appreciably shift the location of 
the false vacuum, which could alter the energy analysis.  The classical equation of 
motion for $\phi$ can be written as
\bea
(-\nabla^2 + M_\phi^2) \phi &\simeq& \langle X^* X \rangle \tilde w' (\phi) \,, 
\eea
where $\tilde w (\phi)$ is the same as $w (\phi)$, but with the $a_2$ term removed (since it is already 
absorbed in the definition of $M_\phi^2$), and $'$ denotes the derivative with respect to $\phi$.  For
the constant solution $\delta \phi$, it is straightforward to see that $\delta \phi \ll v$, 
provided that the value $\phi_{\rm min}$ which locally minimizes 
$\tilde w (\phi)$ satisfies $\phi_{\rm min} \ll v$.  
The correction to the false vacuum energy (and thus the PBH mass) 
is therefore small.  

Note, these considerations have little impact on the true vacuum, for which
$\langle X^* X \rangle \sim 0$.  However, in the true vacuum we have $m_{X,\rm true}^2 \sim w(v)$, 
and we require $m_{X,\rm true}^2 \gg T_*^2 \sim v^2$.  But since $v \gg \Lambda$, it 
is easy to arrange $w(v) \gg v^2$, even if $a_n < 1$.  This is not surprising, since we should 
not expect the mass of $X$ in the true vacuum to be constrained by the low-energy expansion of 
the Yukawa interaction in the false vacuum.

\subsection{Collapse to a PBH}

For the condensate to form a PBH, an additional potential term is required to destabilize the Q-ball.  For simplicity, we consider the case in which 
collapse occurs not because of gravity, but because of a Yukawa potential of the form, 
\bea
V &=& -\frac{g^2}{4\pi r} e^{-M_\phi r} ,
\eea
where $M_\phi$ is the mass in the false vacuum of a heavy particle $\phi$ that mediates the interaction between the condensate constituents $X$.  
In the relativistic limit, for a scalar interaction, $g \sim a_1 \Lambda (R/m_X)^{1/2}$, which can be determined from the $XX \to XX$ scattering cross section via $\phi$ exchange, 
\begin{equation}
\frac{d\sigma}{d\Omega}\propto \frac{1}{E^2} \left(\frac{a_1^2 \Lambda^2}{\vec{q}^2 + M_\phi^2} \right)^2\,,
\end{equation}
where $\vec{q}$ is the three-momentum transfer.
Because $\Lambda > 1/R > m_X$, 
we have $g > a_1$.
Note that $\phi$ is not itself part of the condensate, whose constituents are nearly massless in the false vacuum.

We first consider the case in which $M_\phi R \gg 1$ at the time of condensate formation due to thermal corrections to the effective potential ($M_\phi^2 \propto T^2$).  
Then, the Yukawa interaction is short-ranged with respect to 
the size of the false vacuum bubble.  
The contribution of the Yukawa potential to the free energy of the bubble is then given by~\cite{Kawana:2021tde}
\bea
\langle V \rangle_{\rm bubble} &\propto &  - \frac{g^2 }{4\pi} \frac{(Q/ (M_\phi R)^3 )^2}{M_\phi^{-1}} 
\left(M_\phi R \right)^3 
= -  \frac{g^2 }{4\pi} \frac{Q^2}{M_\phi^2 R^3 }\, .
\eea

The free energy of the false vacuum bubble can then be written as 
\bea
F &=& \frac{\pi}{R }Q
-\frac{2 \pi^3}{135} R^3 T^4 + \frac{4\pi}{3} R^3 U_0 (T) 
-  \frac{a_1^2 c_1}{4\pi} \frac{Q^2 \Lambda^2}{m_X M_\phi^2 R^2 } \,,
\eea
where $c_1$ is an ${\cal O}(1)$ geometric factor.  

For simplicity, we consider the case in which, initially, $M_\phi$ is large enough that 
the Yukawa contribution to the free energy is negligible.  
Thus, as we have seen, a stable Q-ball will form with $R \propto Q^{1/4} [U_0(T)]^{-1/4}$.  As $T$ decreases, $M_\phi$ also decreases, and eventually the 
Yukawa contribution becomes important and may destabilize the Q-ball.

Defining $\tilde U_0 = U_0 - (\pi^2/90) T^4 >0$, we may then write
\bea
\frac{R^3}{4\pi \tilde U_0} \frac{dF}{dR} &\equiv& f(R) =   R^5 - p R
+  q \,, 
\\
p &=& \frac{ Q}{4 \tilde U_0 } \,, 
\nonumber\\
q &=& \frac{a_1^2 c_1}{8\pi^2 \tilde U_0} \frac{Q^2 \Lambda^2}{m_X M_\phi^2} \nonumber\,.
\eea
The false vacuum bubble is stabilized when $f(R) = 0$ (equivalently, $dF/dR = 0$).  Initially, 
$M_\phi$ is large and the Yukawa contribution is negligible.  Then, as $R$ increases from 
zero, $f(R)$ becomes negative, reaches a minimum at $R = (p/5)^{1/4} = [Q / 20 \tilde U_0]^{1/4}$ and then monotonically increases, 
crossing zero at $R = p^{1/4} = [Q / 4 \tilde U_0]^{1/4}$, i.e., the bubble is initially stabilized at $R = [Q / 4 \tilde U_0]^{1/4}$.  As $M_\phi$ decreases and the 
constant term in $f(R)$ increases, the stable bubble radius decreases until it reaches $R = [Q / 20 \tilde U_0]^{1/4}$, 
when $q =(4/5) 5^{-1/4}  p^{5/4} $.  
As $M_\phi$ decreases further, there is no longer a stable solution, and the Q-ball collapses.  
Thus, the Q-ball will be unstable to collapse if 
\bea
M_\phi^2 &<&  \frac{a_1^2 c_1 5^{5/4} }{4\sqrt{2}\pi^2  } \frac{Q^{3/4} \Lambda^2}{ m_X}  \tilde U_0^{1/4}\, .
\eea
As $T$ falls, and the thermal contribution to $M_\phi$ decreases, 
this condition will eventually be satisfied,  and the
collapse of the condensate occurs.

Since $M_\phi \sim T$, we find that the Q-ball collapses to a PBH at temperature 
$T_{\rm PBH}$ given by
\bea
%T_{\textnormal{PBH}} &\propto&  g Q^{1/4} [\tilde U_0(T)]^{1/4} .
T_{\textnormal{PBH}} &\propto&  (a_1 \Lambda  / m_X^{1/2} ) Q^{3/8} \tilde U_0^{1/8} 
\sim g  Q^{1/4} \left[ \tilde U_0 (T) \right]^{1/4}\, .
\label{eq:TPBHprop}
\eea
But the temperature at which the condensate forms is given by $T_{\rm EC} \propto Q^{1/12} [U_0(T)]^{1/4} $.  
Our analysis has assumed the PBH formation temperature is $T_{\rm PBH} \ll [U_0(T)]^{1/4}$, which leads to the coupling 
bound $g < {\cal O}(Q^{-1/4})$ from~Eq.~(\ref{eq:TPBHprop}). 
Note that this bound also implies $T_{\rm PBH} \ll T_{\rm EC}$.
For larger Yukawa couplings, the instability to PBH collapse 
occurs before the formation of a condensate, and possibly out of equilibrium.  In that case, a more detailed numerical calculation is needed, which is beyond the scope of this work.  We focus on the case in which collapse to a PBH occurs after the scalar condensate has formed.

Note that if the Yukawa interaction is negligible, we have $R = p^{1/4}$ in equilibrium, 
and the free energy of the false vacuum bubble is
\bea
F_{q \sim 0} &=& 4 \pi \tilde U_0 (T) \left[\frac{R^3}{3} + \frac{p}{R} \right] =  4 \pi \tilde U_0 (T) 
\left[\frac{4}{3} p^{3/4} \right] .
\eea
But the Q-ball becomes unstable to collapse when $q = 4 (p/5)^{5/4}$.  Just before collapse, 
we have $R = (p/5)^{1/4}$, implying 
\bea
F_\mathrm{collapse} &=& 4 \pi \tilde U_0 (T) \left[\frac{R^3}{3} + \frac{p}{R} - \frac{q}{2 R^2} \right] 
\nonumber\\
 &=& 4 \pi \tilde U_0 (T) \left(\frac{p}{5} \right)^{3/4} \left[\frac{1}{3} + 5 -2 \right] 
= 4 \pi \tilde U_0 (T) 
\left[\frac{4}{3} p^{3/4} \right] \left(\frac{1}{2} 5^{1/4} \right)\,.
\label{eq:collapse}
\eea
Thus, the Yukawa interaction causes the 
free energy of the condensate to decrease by $\sim 25\%$ before it collapses into a PBH.  The 
precise amount will depend in detail on the interaction, and the effect will be small compared to 
other uncertainties.  As such, we neglect this effect.

To find the mass of the PBH, it is necessary to find the energy of the false vacuum bubble at the 
time when it collapses.
We see from Eq.~(\ref{eq:collapse}) that the false vacuum, kinetic and Yukawa contributions to the 
free energy density at the point of collapse are in the proportion $1/3:5:-2$. So $10\%$ of the free energy of the false vacuum bubble arises from the false vacuum free energy 
density itself, while the remaining $90\%$ is due to the kinetic energy and Yukawa interactions of the 
spin-0 particles.  Since most of the scalar particles are in the low-entropy condensed phase, their free energy is the same 
as their energy.  
The entropy of the scalar particles in the normal phase is non-trivial, but we find that the free energy 
of these particles is small compared to that of the false vacuum when a scalar condensate has formed.
The energy density of the false vacuum,
$U_0(T) - T (dU_0/dT)$, is equal to the latent heat density released while the false vacuum converts to the true vacuum as bubbles shrink (since the true vacuum energy density is negligible). 
But since the false vacuum  energy density is itself a small contribution to the 
energy density within the bubble, we approximate the mass of the PBH with the free energy of the false vacuum 
bubble at the time of collapse ($M_{\textnormal{PBH}} \sim F_{\rm collapse} \sim F_{q \sim 0}$), and neglect the energy of the 
scalar particles in the normal phase ($\tilde U_0 \sim U_0$).

Note that collapse to a Q-ball occurs when 
$M_{\phi} \sim g Q^{1/4} U_0^{1/4} \sim g Q^{1/2} /R$, and we have assumed $g < {\cal O}(Q^{-1/4})$.  
Thus, we may take couplings which are large enough that $M_\phi R \gg 1$ at the point of Q-ball collapse to 
a PBH.  But if $g$ (or, equivalently, 
$a_1$) is taken sufficiently small, then collapse may not occur before $M_\phi R \ll 1$.  In this case, 
before the Q-ball collapses, the force will become long-range on the scale of the entire Q-ball, and we will 
find $\langle V \rangle_\mathrm{bubble} \propto g^2 Q^2 /R \propto a_1^2 \Lambda^2 Q^2 /m_X$, which is 
independent of $R$.  For such small values of the coupling, the potential will no longer destabilize the 
Q-ball.

\subsection{Comparison to Fermi-balls }

It is worthwhile to compare these results to the Fermi-ball case.
 %In that case, the details of the phase transition are essentially unchanged.  In particular, 
The radius distribution of false vacuum bubbles, the number of particles per false vacuum bubble ($Q$), and the 
energy scale $[U_0(T)]^{1/4}$ are unchanged,  
yielding the same number density of PBHs.
What changes is the parametric dependence of $M_{\textnormal{PBH}}$ on the only dimensionless quantity, $Q$.  
In particular, 
we have $M_{\textnormal{PBH}}^{\rm boson} \propto Q^{-1/4} M_{\textnormal{PBH}}^{\rm fermion} \ll M_{\textnormal{PBH}}^{\rm fermion}$.  
Thus, for Q-balls 
we get the same PBH number density as for Fermi-balls, but the mass of each PBH is suppressed by a factor of 
roughly $Q^{-1/4}$, giving a much smaller PBH energy density.

This happens because after each false vacuum bubble forms, it shrinks (and releases latent heat) until the point of 
collapse, at which point the energy of the bubble becomes the PBH mass.  In the Fermi-ball case, 
%of a fermion condensate, 
the size of the stable false vacuum bubble scales as $R \propto Q^{1/3}$, which is the same scaling relation as when 
the particles form a classical relativistic gas (the only difference being the scaling of $R$ with $T$ and $U_0 (T)$).
But the radius of a Q-ball is parametrically smaller before it 
finally collapses, yielding a much smaller PBH.

\section{An Example: a Quartic Effective Potential}
\label{sec:quartic}

As a concrete example, consider a thermal effective potential quartic in a single 
real scalar field $\phi$, which also mediates the Yukawa interaction between the $X$:
\bea
U(\phi, T) &=& \Lambda^4 \left[ \left(-\frac{1}{2} + c \left(\frac{T}{v} \right)^2 \right)\left( \frac{\phi}{v} \right)^2
+ b \frac{T}{v} \left( \frac{\phi}{v} \right)^3 + \frac{1}{4} \left( \frac{\phi}{v} \right)^4 \right] + \Lambda_0^4\,,
\eea
 where $b$ and $c$ are dimensionless constants satisfying $c>0$, $b<0$, $c/b^2 > 1$, and 
$\Lambda_0^4$ is chosen so that the vacuum energy density of the true vacuum is given by the cosmological 
constant.  $U(\phi, T=0) $ is minimized at $\phi = v$.

Note that the $b$- and $c$-dependent terms in the thermal effective potential arise after integrating out all the other fields which couple to $\phi$.  Thus, the effective 
potential includes quantum corrections due to $X$ but 
also includes corrections due to any other fields which couple to $\phi$.  Since our goal is not to analyze any 
particular complete ultraviolet model, we do not compute the coefficients $b$ and $c$ from first principles.  Instead, 
we treat them as free parameters which we scan over, under the assumption that any value within our 
scan can be obtained by integrating out some set of additional degrees of freedom, including $X$.  Note also that 
we make no attempt to address fine-tuning issues in the generation of this thermal effective potential from a 
ultraviolet-complete fundamental theory.

For $T > T_c = v/\sqrt{2(c-b^2)}$, the global minimum of the thermal effective potential is at $\phi = 0$.  This is the high-temperature 
phase, which corresponds to the false vacuum.  For $T < T_c$, the global minimum is
\bea
\phi_{\rm true} &=& \frac{-3bT + v\sqrt{9b^2(T/v)^2 + 4(1-2c(T/v)^2)}}{2}\,.
\eea
This is the true vacuum.  Since $U(\phi=0, T)=\Lambda_0^4$,  the difference in free energy density 
between the false vacuum and true vacuum is  
\bea
U_0 (T) &=& \Lambda_0^4 - U(\phi_{\rm true}, T) >0 . 
\eea
Note that since there are no non-gravitational couplings between the dark sector and SM sector, the temperature that appears in 
the thermal effective potential is the temperature of the dark sector, which need not be the same as that of the SM sector.

Below the critical temperature $T_C$, the Euclidean action has a bounce solution~\cite{wainwright2012:cosmotransitions,Akula:2016gpl}, 
which determines the rate of tunneling to the true 
vacuum.  An analytic approximation for the 
Euclidean action is~\cite{Dine:1992wr,Guo:2020grp}
\bea
S_E (T) &=& 4.85 \frac{\Xi^3}{\Sigma^2 T^3} \left[1 + \frac{h}{4} \left(1 + \frac{2.4}{1-h} + \frac{0.26}{(1-h)^2} \right) \right]\,,
\eea
where 
\bea
\Xi^2 &\equiv& \frac{2 \Lambda^4}{v^2} \left(c \left(\frac{T}{v} \right)^2 -\frac{1}{2}\right)\,,
\nonumber\\
\Sigma &\equiv& -b \left( \frac{\Lambda}{v} \right)^4\,,
\nonumber\\
h &\equiv& \frac{1}{2} \left( \frac{\Xi}{\Sigma T} \right)^2 \left( \frac{\Lambda}{v} \right)^4\,.
\eea
The nucleation rate of the true vacuum per volume can then be written as \cite{Linde:1981zj}
\bea
\Gamma_* (T) &\sim& T^4 \exp[-S_E (T)]\,. 
\eea
At the percolation temperature, $S_E$ satisfies~\cite{Megevand:2016lpr}
\bea
S_E (T_*) - \frac{3}{2} \log \frac{S_E (T_*)}{2\pi}   &=& 4 \log \frac{T_*}{H(T_*)} -4 \log [T_* S'_E (T_*)] 
+\log [8\pi v_w^3]\,,
\label{eq:SE_TN}
\eea
where $H(T)$ is the Hubble parameter at temperature $T$.  For simplicity, we take $v_w = 1/\sqrt{3}$.  To determine 
the Hubble parameter, we take the number of relativistic degrees of freedom of the dark sector to be 3.

The latent heat released during the phase transition is then given by 
\bea
\mathcal E &=& U_0 (T_*) - T_* \left. \frac{\partial U_0}{\partial T} \right|_{T=T_*} .
\eea
The strength of the FOPT can be parameterized by
\bea
\alpha &=& \frac{\mathcal E}{\rho_R (T_*)} ,
\eea
where $\rho_R(T)$ is the total radiation energy density of the dark and SM sectors at dark sector temperature $T$; see Section~\ref{secc}. 

\subsection{Relation between PBH mass and percolation temperature}

To illustrate the relationship between the PBH mass and percolation temperature, we vary $v$, while keeping the dimensionless parameters $b$, $c$ and $\Lambda / v$ fixed; $v$ is the only energy scale in the potential.  Thus, we will find $T_* \propto v$, $U_0 (T_*) \propto T_*^4$ up to logarithmic 
corrections induced by the $H(T_*)$ dependence in Eq.~(\ref{eq:SE_TN}), which introduces the Planck mass, $M_{\rm Pl}$, as an additional 
dimensionful quantity.  
In particular, the value of $S_E (T_*)$ which solves Eq.~(\ref{eq:SE_TN}) will only vary slowly 
with $T_*$, but $\exp[-S_E (T_*)] \propto (T_*/M_{\rm Pl})^4$. Then, since $\Gamma_* (T_*) \propto T_*^8 / M_{\rm Pl}^4$, we find 
$Q \propto T_*^{-3}$, and $M_{\textnormal{PBH}} \propto T_*^{-5/4}$.

Thus, for phase transitions with lower percolation temperatures, the comoving false vacuum bubbles
at percolation with be larger, enclosing more particles, and leading to larger PBHs.  Since PBHs arising from Q-balls are already much smaller than those arising from Fermi-balls
it is only at the lowest temperatures that Q-balls will produce a PBH large enough to survive past 
recombination.

\section{Constraints and Observational Signals}
\label{sec:constraints}

We assume that the latent heat released during the FOPT is deposited entirely in the dark sector. This contributes to the
effective number of massless neutrinos ($\Delta N_{\rm eff}$). 
The FOPT may also produce an observable stochastic gravitational wave (GW) signal. Moreover, the PBHs produced from Q-ball collapse 
may produce detectable gamma-ray signals if they survived past the epoch of recombination.

\subsection{Gravitational waves from the first-order phase transition}

The GW signal due to sound waves produced  during the FOPT may be expressed in double power-law form~\cite{Caprini:2024hue}. For  true vacuum bubbles of average radius $R$, at percolation time $t_*$, the amplitude of the GW signal is given by
\bea
\Omega (f) &\simeq& \frac{0.11}{\pi} \left(\sqrt{2} + \frac{2 f_2/f_1}{1+ f_2^2 / f_1^2} \right) 
\left(\frac{a_*}{a_0} \right)^4 \left(\frac{H_*}{H_0}\right)^2 (H_* R)
~{\rm min} \left[2 H_* R \sqrt{\frac{1+\alpha}{1.8\kappa \alpha}} ,1 \right]
%(H_* \tau_{sw}) 
\left(\frac{0.6 \kappa \alpha}{1+\alpha} \right)^2 S(f)\,,
\nonumber\\
S(f) &=& \left(\frac{f}{f_2} \right)^3 \frac{2}{1+ f^4 / f_2^4}  
\frac{1+ f_2^2 / f_1^2}{1+ f^2 / f_1^2}\,,
\label{eq:OmegaGW_Distrib}
\eea
where the frequency scales are given by 
%{\bf There's a factor of 0.6 in $K$ in Ref. 12, which is missing here.}
\bea
f_1 &=& 0.2 \frac{H_{*,0}}{R H_*} = \frac{0.2}{R} \frac{a_*}{a_0}\,,
\nonumber\\
f_2 &=&  0.5 \frac{H_{*,0}}{\Delta_w R H_*} = \frac{0.5 }{\Delta_w R } \frac{a_*}{a_0}\,.
\eea
Here, $H_0= 100h$~km/s/Mpc, $H_{*,0} = H_* (a_*/a_0)$, and $\Delta_w = (|v_w - c_s|)/{\rm max}[v_w,c_s] \lesssim 1$, 
where $c_s$ is the sound speed. $\kappa$ is an efficiency factor related to how efficiently the latent heat is converted into sound waves.  
For a phase transition in the dark sector, $\kappa$ can be 
${\cal O}(1)$~\cite{Nakai:2020oit}. We make the optimistic assumptions that $\Delta_w =1$ and $\kappa = 1$.
If $\Delta_w \ll 1$, then the 
gravitational wave signal will be suppressed by a factor $\Delta_w^{-2}$. The factors of $a_* / a_0$ arise from redshifting the signal from the time the percolation time until today.  This factor is determined by the temperature of 
the SM sector at the time of percolation, which may be different from $T_*$.

The distribution of the true vacuum bubble radius at percolation is given by \cite{Marfatia:2024cac}
\bea
\frac{dn_{\rm TV}}{dR} &=& \frac{3}{4\pi R_{f*}^4} x^{-1} \exp[-1/x]\,,
\eea
where $x = \exp[6^{1/4} R / R_{f*}]$.  The gravitational wave signal is then obtained by 
weighting Eq.~(\ref{eq:OmegaGW_Distrib}), by 
$dn_{\rm TV}/dR$, as in Ref.~\cite{Marfatia:2024cac}.

\subsection{Gamma-rays from Hawking evaporation}

The Hawking evaporation of PBHs can produce detectable gamma rays, providing constraints on 
the PBH mass distribution~\cite{Auffinger:2022khh}.
Bounds on primordial black holes are presented under the assumption of a monochromatic 
mass distribution.  In the monochromatic case, the fraction of the total energy density that consists of mass $M$
PBHs at the time of their formation is given by \cite{Carr:2020gox} 
\bea
\beta (M) &=& \frac{M  n_{\textnormal{PBH}} (t_i) }{ \rho_{\rm tot} (t_i)} =   \frac{M}{ \rho_{\rm tot} (t_i)} \frac{F(T_*)}{(4\pi/3) R_f^3}\,,
\eea
where $t_i$ is the time of black hole formation. 
It is common in the literature for constraints to instead be 
expressed in terms of $\beta' (M)$, defined as 
\bea
\beta' (M) &=& \gamma^{1/2} \left(\frac{g_\rho}{106.75} \right)^{-1/4}  \left( \frac{h}{0.67} \right)^2 \beta (M)\,,
\label{eq:betaprime}
\eea
where 
\bea
\gamma^{1/2} &\equiv& \left( \frac{M}{c^3 t_i G_{\rm N}} \right)^{1/2} = 4.48 \times 10^{-12} \left(\frac{g_\rho}{106.75} \right)^{1/4} 
\frac{T_*}{\mev} \left(\frac{M}{10^{-18} M_\odot} \right)^{1/2}\,.
\eea
The fraction of dark matter comprised of PBHs can then be expressed as~\cite{Carr:2020gox} 
\bea
f_{\textnormal{PBH}} &=& 3.81 \times 10^{17} \left(\frac{M}{10^{-18} M_\odot} \right)^{-1/2} \beta' (M)\,, 
\nonumber\\
&=& 1.7 \times 10^6 \left( \frac{h}{0.67} \right)^2 \frac{T_*}{\mev} ~ \beta (M)\,.
\eea

However, the PBH mass distribution is not expected to be monochromatic.  For PBHs that survive past 
recombination ($M_{\textnormal{PBH}} \gtrsim 10^{15}\g$), observational constraints on $\beta' (M)$ tend to weaken with increasing $M$.  
As such, even a small density of light PBHs can yield tight constraints.
We can set conservative observational bounds on PBHs that survive past recombination by defining
\bea
\tilde \beta (M) &=& \frac{M}{\rho_{\rm tot}(t_i)}\int_{10^{15}\g}^M dM' ~ \frac{dn_{\textnormal{PBH}}}{dM'}\,,\quad \quad \frac{dn_{\rm PBH}}{dM'} = \frac{dn_{\rm FV}}{dR}\frac{dR}{dQ}\frac{dQ}{dM'}\,,
\eea
where $dn_{\rm FV}/dR$, $dR/dQ$, and $dQ/dM'$ are obtained from Eqs.~(\ref{eq:dndr}), (\ref{eq:QofR}), and (\ref{eq:MofQ}), respectively, and we have used the fact that every false vacuum bubble produces a PBH. $\tilde \beta (M)$ is the same as $\beta (M)$ if all PBHs with mass $M_{\textnormal{PBH}} > M$ are ignored, and all PBHs with mass in the range 
$10^{15}\g< M_{\textnormal{PBH}}<M$ instead have $M_{\textnormal{PBH}} =M$ (since constraints are weaker for heavier PBHs).  
One can then compute $\tilde \beta' (M)$ from $\tilde \beta (M)$ using Eq.~(\ref{eq:betaprime}).
For a given model, if $\tilde \beta' (M)$ lies in the excluded region of $\beta' (M)$ for any choice of $M$, then the model is excluded. 

For the models we consider, it turns out that $\tilde \beta' (M)$ grows with $M$ with roughly the same slope as the 
$\beta' (M)$ exclusion contours.  Thus, in determining the sensitivity of gamma-ray observatories is sufficient to 
evaluate $\beta' (M)$ assuming a monochromatic distribution of PBH masses arising from collapse of a false vacuum bubble with fixed size 
$R_{f*}$.

\subsection{Gamma-rays from superradiance}

PBHs may acquire spin from progenitor false vacuum bubbles if they have angular momentum~\cite{Acuna:2025vdf}.
 The $X$ particles (and their antiparticles) can be regenerated in the vicinity of 
the rotating PBH through superradiance, and their subsequent decays can provide another observable signal~\cite{Brito:2015oca}. For a rotating PBH with angular velocity $\omega_H$ (as seen by an observer at infinity) at the outer event horizon, that satisfies 
$\omega_H > \textnormal{Re}(\omega)/\mathtt m$ for a particle with (complex) angular frequency $\omega$, and azimuthal number $\mathtt m$, the PBH will exhibit a superradiant instability. This instability, which extracts energy and angular momentum from the PBH, can lead to the exponential growth of the $X$ field\footnote{The real part of the angular frequency is related to the particle mass via $\omega = m_X(1-\alpha^2/(2n^2))$, to $\mathcal{O}(\alpha^2)$ in the dimensionless coupling $\alpha = m_X G_{\rm N} M_{\rm PBH}$~\cite{Dolan:2007mj,Brito:2015oca,Baumann:2018vus}.} creating a bosonic cloud around the PBH populated by a large number of particles. In the absence of self-interactions this number is $N_X = G_{\rm N} M_{\textnormal{PBH}}^2\Delta a_*/\mathtt m$, where $\Delta a_*$ is the change in the dimensionless spin of the PBH, $a_* = J/(G_{\rm N} M_{\rm PBH}^2)$, where $J$ is the PBH's angular momentum) induced by the superradiant cloud. 
A superradiated scalar with a coupling to the electromagnetic field can decay to two photons, each of energy $E_{\gamma} = m_X/2$, where the strength of the coupling can be related to the scalar self-interaction.
This decay from the superradiated cloud can thus produce an additional PBH signature, though ontological parsimony 
is sacrificed (sufficient PBH spin and a boson mass in the proper range are needed). We are not adopting a specific mechanism that could produce a large dimensionless spin (and to our knowledge, there are no detailed calculations of PBH spins formed in a phase transition), rather, we take a phenomenological approach to explore the possible signatures from superradiance in the PBH from FOPT scenario if large spins can be achieved.

\subsection{$\Delta N_{\rm eff}$}
\label{secc}

Because the dark sector and 
SM sector are decoupled, they need not be at the same temperature. 
We define the ratio of temperatures $\zeta (T) \equiv T_{\rm SM}/T$, where $T_{\rm SM}$ is the temperature of the SM sector when the dark sector is at temperature $T$. At percolation, $\zeta_* = T_{\rm SM*} / T_*$.  
The latent heat released during the phase transition will heat the dark sector of the true vacuum. The temperature of the bath in the true vacuum after the phase transition $T'$ satisfies
\bea
g_{\rho D}(T') \frac{\pi^2}{30} T^{'4} &=& g_{\rho D}(T_*) \frac{\pi^2}{30} T_*^{4} + \mathcal E\,, 
\eea
where $g_{\rho D}$ is the number of dark sector relativistic degrees of freedom.
The ratio of the SM temperature to the dark sector temperature when the dark sector temperature is $T'$ is then given by
\bea
\zeta (T') = \zeta_* \times \frac{T_*}{T'} \times \left(\frac{g_{s{\rm{SM}}}(T_{\rm SM*})}{g_{s\rm{SM}}(T'_{\rm SM})} \right)^{1/3}\,,
\eea
where $g_{s{\rm SM}}$ is the number of SM entropy degrees of freedom. We then have
\bea
\rho_R (T') &=& \left[g_{\rho {\rm SM}} (T'_{\rm{SM}}) + g_{\rho D} (T')~ \zeta(T')^{-4}  \right] \frac{\pi^2}{30} T_{\rm SM}^{'4} \,,
\nonumber\\
\Delta N_{\rm eff} (T') &=& \frac{4}{7} \left(\frac{2}{11} g_{s {\rm SM}}(T'_{\rm SM}) \right)^{-4/3} \zeta(T')^{-4} ~g_{\rho D} (T')\,.
\eea
If $\phi$ is relativistic below $T_c$, its contribution to $N_{\rm eff}$ is included via $g_{\rho D}$. 
On the other hand, if $\phi$ is nonrelativistic after the FOPT, it may behave as dark matter. A mechanism to avoid overclosing the Universe that keeps the value of $g_{\rho D}$ unchanged can be found in Ref.~\cite{Marfatia:2022jiz}.

\section{Results}
\label{sec:results}

For models that satisfy current constraints and can produce multimessenger signals 
at future observatories, the percolation temperature is restricted to the range $100~\kev \lesssim T_* \lesssim 1~\mev$ for $\eta_X = 1$.
To understand the reason for this qualitatively, hold the dimensionless parameters $b$, $c$, $\Lambda / v$ and $\zeta_*$ fixed, while varying $T_* \propto v$.
As we have seen, the PBH mass decreases with percolation temperature as $M_{\textnormal{PBH}} \propto T_*^{-5/4}$.  
Also, since the radius of the false vacuum bubble scales as $R_{f*} \propto T_*^{-2}$, 
we find $\tilde \beta' (M) \propto M_{\textnormal{PBH}}^{3/2} T_*^{3} \propto T_*^{9/8}$.\footnote{Note that this relation cannot 
be used determine the scaling of $\tilde \beta'$ with $M_{\textnormal{PBH}}$ or $T_*$ if the dimensionless parameters are varied, because 
$\tilde \beta'$, $M_{\textnormal{PBH}}$ and $T_*$ each have a non-trivial dependence on the dimensionless parameters.}  For 
$T_* \lesssim 100~\kev$, we find that $\tilde \beta'(M)$ is so small that no upcoming observatories will be sensitive 
to the PBH population.\footnote{We find, for example, that for $T_* = 10$~keV, a GW signal with  $\Omega_{\rm GW}h^2 \gtrsim 10^{-16}$ can be generated, but with $f_{\textnormal{PBH}} \lesssim 10^{-10}$, putting it out of reach for planned MeV observatories such as AMEGO-X.}
On the other hand, for $T_* \gtrsim 1~\mev$, $\tilde \beta'$ is relatively large, while the PBH mass is relatively small and in the range where 
constraints on $\tilde \beta'$ are tight, ruling out these scenarios. 

Focusing on this region of parameter space, we find that the GW signal peaks in roughly the $1-100~\nhz$ range, as the frequency scale is set by the percolation temperature.  
The FOPT releases latent heat into the dark sector,
which contributes to $\Delta N_{\rm eff}$. This contribution can be suppressed by setting $\zeta_* >1$, thereby diluting the energy density of dark radiation.  However, this
suppresses the GW signal as well.  
These competing effects shrink the parameter space in which observable GW signals can be found.
 For scenarios with $\Omega_{\rm GW} h^2 \gtrsim 10^{-16}$, which is the estimated sensitivity in the nanohertz range for the
proposed observatories SKA~\cite{Janssen:2014dka} and THEIA~\cite{Garcia-Bellido:2021zgu}, and $\Delta N_{\rm eff} < 0.5$, 
we find $\zeta_* \sim 3\text{--}8$.  

\begin{figure*}[t]
    \centering
    \includegraphics[width=0.8\textwidth]{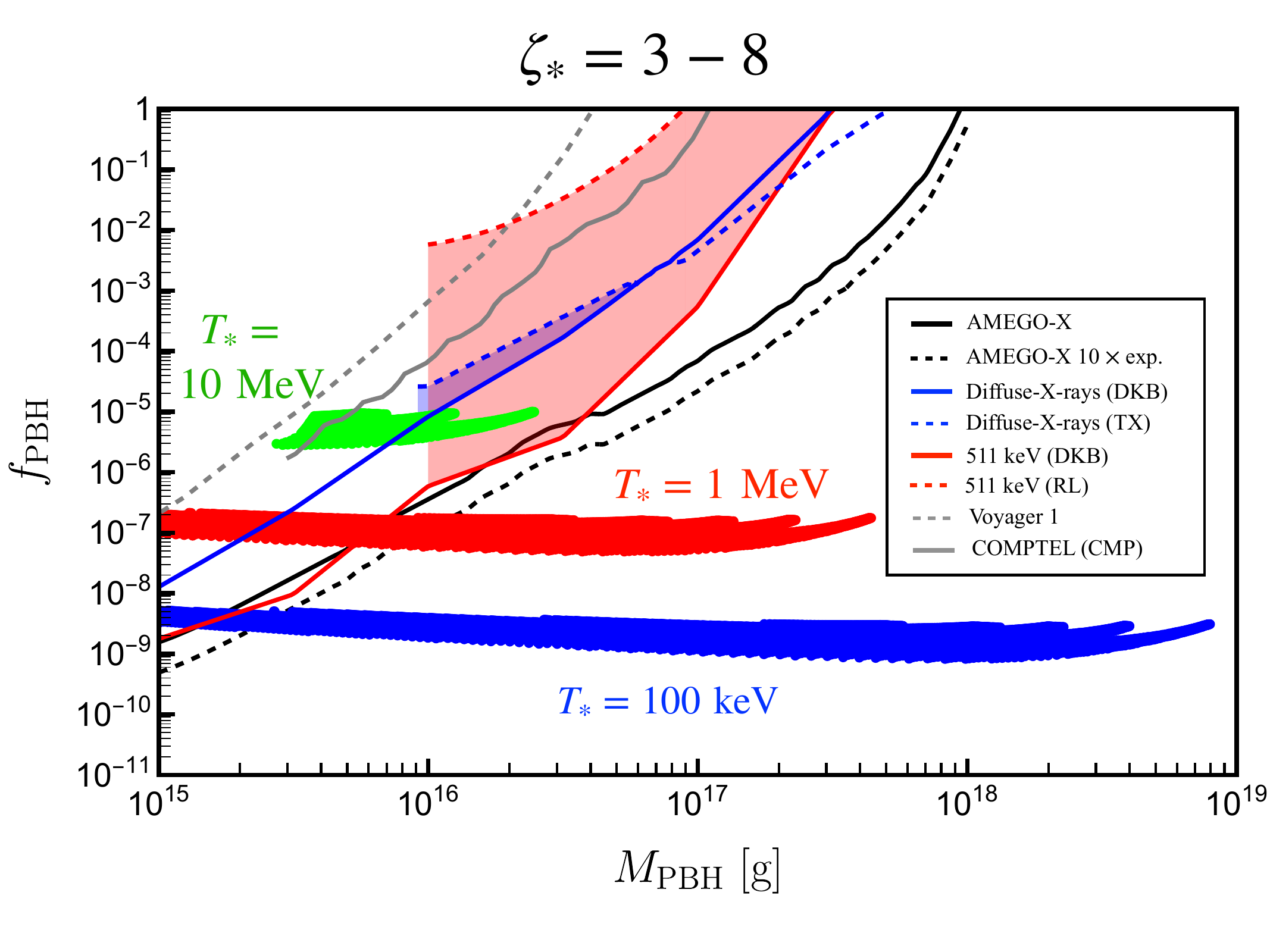}\\
    \caption{Results of a model scan in the $(M_{\textnormal{PBH}}, f_{\textnormal{PBH}})$ plane, for 
    %$\zeta_* =3$, 
    $\Lambda / v = 0.8$ and 
    $T_{*} = 100\kev$ (blue points), $1\mev$ (red points), or 10~MeV (green points).  $\zeta_*$ takes values 
    between $3$ and $8$.  
    For all models, $M_{\rm PBH} > 10^{15}\g$ 
    and $h^2 \Omega_{\rm GW}^{\rm max} > 10^{-16}$, with an optimistic efficiency factor $\kappa = 1$.
    Current constraints on the Hawking evaporation of PBHs from VOYAGER 1 $e^{\pm}$ (gray dashed line)~\cite{Boudaud:2018hqb}, COMPTEL gamma-ray (gray line~\cite{Coogan:2020tuf}) diffuse X-ray (blue lines (DKB)~\cite{DelaTorreLuque:2024qms} and (TX)~\cite{Tan:2024nbx}) and $511~\kev$ line searches 
    (red dashed line (RL)~\cite{Laha:2019ssq} and the recent update in solid red~\cite{DelaTorreLuque:2024qms}) are also shown, along with the projected sensitivity of AMEGO-X (solid black line) and AMEGO-X with $10\times$ nominal 
    exposure (black dashed line).  The lightly-shaded red and blue regions indicate that the stronger solid-line limits (relative to the previous dashed-line limits) placed by the 511~keV line and diffuse X-ray data~(DKB) have uncertainties due to cosmic-ray propagation modeling.
    }
    \label{fig:ConstraintsSensitivity}
\end{figure*}

To illustrate these results, 
we scan over the parameters $b$, $c$ and $\zeta_*$  
for $\Lambda/v = 0.8$ and  $T_* = 100~\kev$ (blue) and $1~\mev$ (red) (having fixed $T_*$, one can solve for $v$).    In Fig.~\ref{fig:ConstraintsSensitivity}, we plot the values of $M_{\textnormal{PBH}}$ and $f_{\textnormal{PBH}}$  obtained from this scan, 
assuming that, at the time of percolation, false vacuum bubbles all have radius $R_{f*}$.  We also plot current constraints 
on $f_{\textnormal{PBH}}$, and the future sensitivity of AMEGO-X (including an extrapolation of the sensitivity to $10\times $ the exposure).  
We only plot points for which $M_{\textnormal{PBH}} > 10^{15}\g$ (lighter PBHs that satisfy observational constraints evaporate away 
before the current epoch), $\Delta N_{\rm eff} \leq 0.5$, and  the maximum amplitude of the GW signal satisfies $h^2 \Omega_{\rm GW}^{\rm max} \geq 10^{-16}$. Note that $h^2 \Omega_{\rm GW}^{\rm max}$
scales as a large power of $\Lambda / v$~\cite{Dent:2022bcd}.  Although $\Lambda / v \lesssim 1$ in order for the quartic coupling 
to be perturbative, if $\Lambda / v \ll 1$, then the GW signal will be too small to be observable at upcoming experiments.  These considerations, as well as computational efficiency, motivate our choice to fix $\Lambda / v =0.8$.
The largest gravitational wave signal found is $h^2 \Omega_{\rm GW}^{\rm max} \sim 10^{-14}$. For these results, we have taken the most optimistic value for the conversion efficiency $\kappa = 1$. Since $\Omega_{\rm GW}\propto \kappa^2$ (see Eq.~\ref{eq:OmegaGW_Distrib}), lowering this value can quickly diminish the sensitivity reach.

For the points in Fig.~\ref{fig:ConstraintsSensitivity}, we find $Q\lesssim 10^{61}$.  To ensure that 
collapse to a PBH occurs only after the condensate has formed, we must then have $g \lesssim {\cal O}(10^{-15})$.  
Note that for $T \lesssim \mev$, these Yukawa couplings 
are still much larger than gravitational couplings.

\begin{figure*}[t]
    \centering
    
    \includegraphics[width=0.8\textwidth]{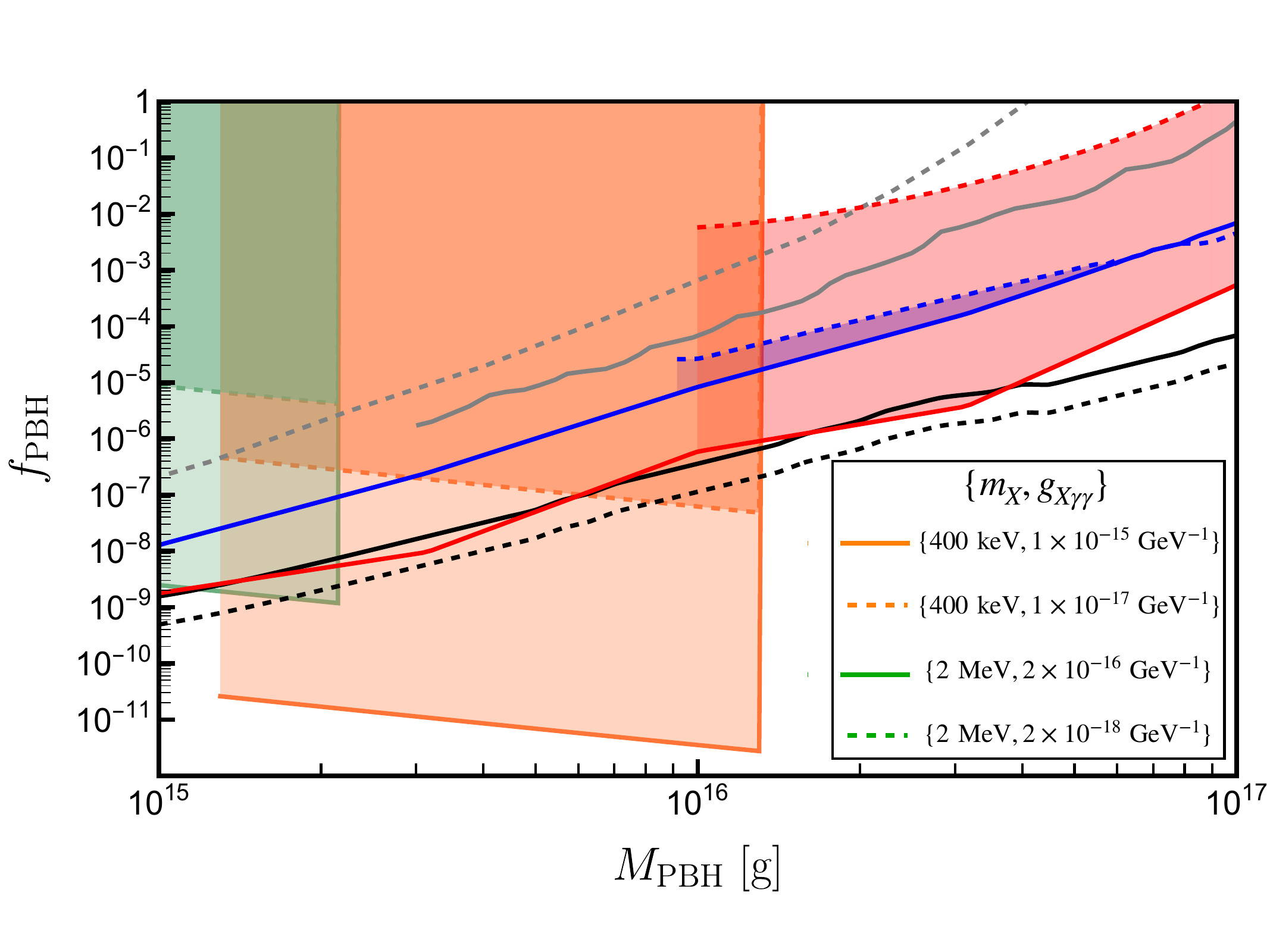}\\[-2.0em]
    \caption{Sensitivity in the $(M_{\textnormal{PBH}}, f_{\textnormal{PBH}})$ plane of AMEGO-X to the decay of a $400~\kev$ (green shaded region) or $2~\mev$ (orange shaded region) scalar $X$ produced 
    through a superradiant instability, assuming $a_* = 0.99$, with $g_{X \gamma \gamma}$ as labeled. These curves are displayed for such a near-extremal spin to show what a possible upper-limit sensitivity reach in this parameter space may be. The other constraints shown are the same as in Fig.~\ref{fig:ConstraintsSensitivity}.}
    \label{fig:superradiance}
\end{figure*}

For a PBH with near-maximal dimensionless spin parameter ($a_* \sim 1$), which we assume as a phenomenological practice (rather than from a first principles calculation) in order to determine the scope of possible PBH signatures,  we have $\omega_H \sim (2G_{\rm N} M_{\textnormal{PBH}})^{-1}$, implying that a superradiant instability only 
occurs for $M_{\textnormal{PBH}} \lesssim (2 G_{\rm N} m_X)^{-1}$.  We find that for $m_X \gtrsim 400~\kev$, the process $X \rightarrow \gamma \gamma$ can yield 
photons energetic enough to be seen at AMEGO-X, and appreciable superradiant $X$ production will occur for $M_{\textnormal{PBH}} \lesssim 1.3 \times 10^{16}~\g$. 
Given $f_{\textnormal{PBH}}$, $N_X$ and $\Gamma_{X \rightarrow \gamma \gamma}$, we estimate the 
flux of photons produced by decay of the superradiantly-produced $X$, and compare to the sensitivity of AMEGO-X to a line at $E_\gamma = m_X/2$.
Although $X$ need not be an axion-like particle, to retain consistency with the literature, we parameterize $\Gamma_{X \rightarrow \gamma \gamma}$ with the 
coupling $g_X\gamma \gamma$ using the relation $\Gamma_{X \rightarrow \gamma \gamma} = g_{X\gamma \gamma}^2 m_X^3 / 64 \pi$.
In Fig.~\ref{fig:superradiance}, we present the sensitivity of AMEGO-X in the $(M_{\textnormal{PBH}}, f_{\textnormal{PBH}})$ plane to $X$ decay after superradiant production, 
for $m_X =400~\kev$ and $2~\mev$.  For this purpose, we have conservatively assumed that the number of superradiantly-produced bosons is 
not saturated by self-interactions; although interactions that couple the scalar to photons will generally induce scalar self-interactions, the relationship 
between them depends on the details of ultraviolet physics, and the self-interaction strength could be small in some models.

For $m_X = 400~\kev\, (2~\mev)$, $g_{X\gamma\gamma}$ can take values for which the sensitivity of AMEGO-X to superradiance signals exceeds 
that of Hawking evaporation, and in particular can probe scenarios in which the PBHs are generated in an FOPT that 
percolates at $T_* = 100~\kev\, (1~\mev)$.  Note that in both these examples we have taken $m_X \gtrsim T_*$, consistent with the fact that $m_X$ in the true vacuum 
should be large enough that $X$ remains confined to the false vacuum. 
However, we have assumed that there is negligible suppression of superradiance due to scalar self-interaction.  If 
the self-interaction is not negligible, then superradiant scalar production (and the associated signal) can be suppressed.

\section{Discussion}
\label{sec:discussion}

The phenomenology of PBH formation from the collapse of a Q-ball  
is very different from that of Fermi-ball collapse.  In particular, it is not possible to obtain 
a value of $f_{\textnormal{PBH}}$ close to unity because Q-balls shrink to a much smaller size than Fermi-balls 
before collapsing.  As a result, the correlation between the gravitational wave and gamma-ray signals is very different from that for the collapse of a fermion condensate, so this correlated signal would be a smoking gun for scalar condensate collapse.  Moreover, the appearance of a photon line-signal produced by the 
superradiant production and subsequent decay of $X$ after PBH formation would be a unique signal, since it requires that the new 
light particle be a boson.

However, obtaining an observable GW signal at planned experiments requires optimistic 
assumptions about the efficiency with which latent heat is converted into sound waves ($\kappa = 1$).  For less efficient 
conversion, observing a multimessenger signal requires improvements in sensitivity in the nanohertz range. 

We also see that scenarios with an observable GW signal tend to have a slightly 
large value of $\Delta N_{\rm eff}$.  Relatively large values of $\Delta N_{\rm eff}\sim 0.4$ could play a 
role in resolving the Hubble tension.  However, such scenarios can also be constrained by CMB-S4 data. Thus,  
CMB, gamma-ray and gravitational wave observatories can play a role in testing this scenario.

Because dark sector radiation can constitute a sizable fraction of the energy density after the phase transition, another interesting possibility arises.  
If the lightest dark sector particle is a cosmologically stable particle with mass $\sim {\cal O}(\ev)$, then at the epoch of matter-radiation equality, this particle will begin to redshift like matter, with a density that is a non-negligibile fraction of the  total matter density.  The particle is therefore a hot dark matter candidate, with a sharp prediction for its mass consistent with current constraints from cosmological observables~\cite{Xu:2021rwg,Peters:2023asu}.

\section{Conclusion}
\label{sec:conclusion}

We studied the formation of primordial black holes via the collapse of Q-balls produced in a first-order phase transition. This scenario provides a new mechanism for seeding 
PBH formation, and leads to correlated signals from the phase transition and from the PBHs themselves. Similar
correlated signals are seen from the collapse of Fermi-balls to PBHs, but the Q-ball 
case has a very different phenomenology, because the PBHs tend to be much smaller.

Although the scenario is general, we considered a specific framework in which the dark FOPT is described by a quartic 
thermal effective potential.  We find that because the PBHs tend to be small, there is a narrow window of parameter space in which 
multi-messenger signal are possible.  In particular, if the percolation temperature is in the $\sim 100~\kev - 1~\mev$ range, 
then the FOPT can yield gravitational waves in the $1-100~\nhz$ range which may be observable at planned observatories.  
The Q-balls can collapse to form $\sim 10^{15}\g$ mass PBHs, 
whose Hawking evaporation can yield signals detectable by future MeV-range gamma-ray telescopes.  
The scalar can also be emitted after PBH formation via a superradiant instability near the PBH, leading 
to decay signals at future MeV-range gamma-ray telescopes.
Moreover, in scenarios in which the gravitational 
wave signal is detectable, dark radiation is also produced, which can alleviate the Hubble tension, yield observable signals
at CMB-S4, and potentially include a hot dark matter candidate.

We have limited our study to the regime of a scalar self-interaction in which the scalars condense before collapsing into a PBH. For stronger self-interactions, collapse may happen in the intermediate regime, when the fraction of particles in the condensed state 
is sizable, but not dominant.  An analysis of this scenario cannot be done analytically but requires a detailed numerical calculation, which would be an interesting topic of future work.

\section*{Acknowledgements}

We thank Pearl Sandick and Tao Xu for useful discussions.
JBD is supported by the National Science Foundation
under grant no. PHY-2412995. BD is supported in part by DOE grant DE-SC0010813. 
JK and DM are supported in part by DOE grant DE-SC0010504. JBD thanks the Mitchell Institute at Texas A\&M University for its hospitality where part of this work was completed. 
The authors acknowledge the Center for Theoretical Underground Physics and Related Areas (CETUP*), the Institute for Underground Science at Sanford Underground Research Facility (SURF), and the South Dakota Science and Technology Authority for hospitality and financial support, as well as for providing a stimulating environment where part of this work was completed.

\newpage
\bibliography{pbhfopt}

@article{Akula:2016gpl,
    author = "Akula, Sujeet and Bal\'azs, Csaba and White, Graham A.",
    title = "{Semi-analytic techniques for calculating bubble wall profiles}",
    eprint = "1608.00008",
    archivePrefix = "arXiv",
    primaryClass = "hep-ph",
    reportNumber = "COEPP-MN-16-19",
    doi = "10.1140/epjc/s10052-016-4519-5",
    journal = "Eur. Phys. J. C",
    volume = "76",
    number = "12",
    pages = "681",
    year = "2016"
}

@article{Lu:2022paj,
    author = "Lu, Philip and Kawana, Kiyoharu and Xie, Ke-Pan",
    title = "{Old phase remnants in first-order phase transitions}",
    eprint = "2202.03439",
    archivePrefix = "arXiv",
    primaryClass = "astro-ph.CO",
    doi = "10.1103/PhysRevD.105.123503",
    journal = "Phys. Rev. D",
    volume = "105",
    number = "12",
    pages = "123503",
    year = "2022"
}

@article{Acuna:2025vdf,
    author = "Acu{\~n}a, Jan Tristram and Marfatia, Danny and Tseng, Po-Yan",
    title = "{Angular momentum of vacuum bubbles in a first-order phase transition}",
    eprint = "2505.09202",
    archivePrefix = "arXiv",
    primaryClass = "hep-ph",
    doi = "10.1088/1475-7516/2026/06/042",
    journal = "JCAP",
    volume = "06",
    pages = "042",
    year = "2026"
}

@article{Megevand:2016lpr,
    author = "Megevand, Ariel and Ramirez, Santiago",
    title = "{Bubble nucleation and growth in very strong cosmological phase transitions}",
    eprint = "1611.05853",
    archivePrefix = "arXiv",
    primaryClass = "astro-ph.CO",
    doi = "10.1016/j.nuclphysb.2017.03.009",
    journal = "Nucl. Phys. B",
    volume = "919",
    pages = "74--109",
    year = "2017"
}

@article{wainwright2012:cosmotransitions,
   title={CosmoTransitions: Computing cosmological phase transition temperatures and bubble profiles with multiple fields},
   volume={183},
   ISSN={0010-4655},
   url={http://dx.doi.org/10.1016/j.cpc.2012.04.004},
   DOI={10.1016/j.cpc.2012.04.004},
   number={9},
   journal={Computer Physics Communications},
   publisher={Elsevier BV},
   author={Wainwright, Carroll L.},
   year={2012},
   month={Sep},
   pages={2006–2013}
}

@article{Dine:1992wr,
    author = "Dine, Michael and Leigh, Robert G. and Huet, Patrick Y. and Linde, Andrei D. and Linde, Dmitri A.",
    title = "{Towards the theory of the electroweak phase transition}",
    eprint = "hep-ph/9203203",
    archivePrefix = "arXiv",
    reportNumber = "SLAC-PUB-5741, SCIPP-92-07, SU-ITP-92-7",
    doi = "10.1103/PhysRevD.46.550",
    journal = "Phys. Rev. D",
    volume = "46",
    pages = "550--571",
    year = "1992"
}

@article{Guo:2020grp,
    author = "Guo, Huai-Ke and Sinha, Kuver and Vagie, Daniel and White, Graham",
    title = "{Phase Transitions in an Expanding Universe: Stochastic Gravitational Waves in Standard and Non-Standard Histories}",
    eprint = "2007.08537",
    archivePrefix = "arXiv",
    primaryClass = "hep-ph",
    doi = "10.1088/1475-7516/2021/01/001",
    journal = "JCAP",
    volume = "01",
    pages = "001",
    year = "2021"
}

@article{Liu:2021svg,
    author = "Liu, Jing and Bian, Ligong and Cai, Rong-Gen and Guo, Zong-Kuan and Wang, Shao-Jiang",
    title = "{Primordial black hole production during first-order phase transitions}",
    eprint = "2106.05637",
    archivePrefix = "arXiv",
    primaryClass = "astro-ph.CO",
    doi = "10.1103/PhysRevD.105.L021303",
    journal = "Phys. Rev. D",
    volume = "105",
    number = "2",
    pages = "L021303",
    year = "2022"
}

@article{Kawana:2021tde,
    author = "Kawana, Kiyoharu and Xie, Ke-Pan",
    title = "{Primordial black holes from a cosmic phase transition: The collapse of Fermi-balls}",
    eprint = "2106.00111",
    archivePrefix = "arXiv",
    primaryClass = "astro-ph.CO",
    doi = "10.1016/j.physletb.2021.136791",
    journal = "Phys. Lett. B",
    volume = "824",
    pages = "136791",
    year = "2022"
}

@article{Hong:2020est,
    author = "Hong, Jeong-Pyong and Jung, Sunghoon and Xie, Ke-Pan",
    title = "{Fermi-ball dark matter from a first-order phase transition}",
    eprint = "2008.04430",
    archivePrefix = "arXiv",
    primaryClass = "hep-ph",
    doi = "10.1103/PhysRevD.102.075028",
    journal = "Phys. Rev. D",
    volume = "102",
    number = "7",
    pages = "075028",
    year = "2020"
}

@article{Marfatia:2021hcp,
    author = "Marfatia, Danny and Tseng, Po-Yan",
    title = "{Correlated signals of first-order phase transitions and primordial black hole evaporation}",
    eprint = "2112.14588",
    archivePrefix = "arXiv",
    primaryClass = "hep-ph",
    doi = "10.1007/JHEP08(2022)001",
    journal = "JHEP",
    volume = "08",
    pages = "001",
    year = "2022",
    note = "[Erratum: JHEP 08, 249 (2022)]"
}

@article{Dent:2022bcd,
    author = "Dent, James B. and Dutta, Bhaskar and Ghosh, Sumit and Kumar, Jason and Runburg, Jack",
    title = "{Sensitivity to dark sector scales from gravitational wave signatures}",
    eprint = "2203.11736",
    archivePrefix = "arXiv",
    primaryClass = "hep-ph",
    reportNumber = "MI-HET-774, KIAS-P22016",
    doi = "10.1007/JHEP08(2022)300",
    journal = "JHEP",
    volume = "08",
    pages = "300",
    year = "2022"
}

@article{Carr:2020gox,
    author = "Carr, Bernard and Kohri, Kazunori and Sendouda, Yuuiti and Yokoyama, Jun'ichi",
    title = "{Constraints on primordial black holes}",
    eprint = "2002.12778",
    archivePrefix = "arXiv",
    primaryClass = "astro-ph.CO",
    reportNumber = "RESCEU-03/20; KEK-Cosmo-249; KEK-TH-2199; IPMU20-0024",
    doi = "10.1088/1361-6633/ac1e31",
    journal = "Rept. Prog. Phys.",
    volume = "84",
    number = "11",
    pages = "116902",
    year = "2021"
}

@article{Coogan:2020tuf,
    author = "Coogan, Adam and Morrison, Logan and Profumo, Stefano",
    title = "{Direct Detection of Hawking Radiation from Asteroid-Mass Primordial Black Holes}",
    eprint = "2010.04797",
    archivePrefix = "arXiv",
    primaryClass = "astro-ph.CO",
    doi = "10.1103/PhysRevLett.126.171101",
    journal = "Phys. Rev. Lett.",
    volume = "126",
    number = "17",
    pages = "171101",
    year = "2021"
}

@article{Auffinger:2022khh,
    author = "Auffinger, J\'er\'emy",
    title = "{Primordial black hole constraints with Hawking radiation\textemdash{}A review}",
    eprint = "2206.02672",
    archivePrefix = "arXiv",
    primaryClass = "astro-ph.CO",
    doi = "10.1016/j.ppnp.2023.104040",
    journal = "Prog. Part. Nucl. Phys.",
    volume = "131",
    pages = "104040",
    year = "2023"
}

@article{Marfatia:2022jiz,
    author = "Marfatia, Danny and Tseng, Po-Yan",
    title = "{Boosted dark matter from primordial black holes produced in a first-order phase transition}",
    eprint = "2212.13035",
    archivePrefix = "arXiv",
    primaryClass = "hep-ph",
    doi = "10.1007/JHEP04(2023)006",
    journal = "JHEP",
    volume = "04",
    pages = "006",
    year = "2023"
}

@article{Marfatia:2024cac,
    author = "Marfatia, Danny and Tseng, Po-Yan and Yeh, Yu-Min",
    title = "{Phenomenology of bubble size distributions in a first-order phase transition}",
    eprint = "2407.15419",
    archivePrefix = "arXiv",
    primaryClass = "hep-ph",
    doi = "10.1140/epjc/s10052-025-14847-x",
    journal = "Eur. Phys. J. C",
    volume = "85",
    number = "10",
    pages = "1150",
    year = "2025"
}

@article{Nakai:2020oit,
    author = "Nakai, Yuichiro and Suzuki, Motoo and Takahashi, Fuminobu and Yamada, Masaki",
    title = "{Gravitational Waves and Dark Radiation from Dark Phase Transition: Connecting NANOGrav Pulsar Timing Data and Hubble Tension}",
    eprint = "2009.09754",
    archivePrefix = "arXiv",
    primaryClass = "astro-ph.CO",
    reportNumber = "TU-1109; IPMU20-0100",
    doi = "10.1016/j.physletb.2021.136238",
    journal = "Phys. Lett. B",
    volume = "816",
    pages = "136238",
    year = "2021"
}

@article{Xu:2021rwg,
    author = "Xu, Weishuang Linda and Mu\~noz, Julian B. and Dvorkin, Cora",
    title = "{Cosmological constraints on light but massive relics}",
    eprint = "2107.09664",
    archivePrefix = "arXiv",
    primaryClass = "astro-ph.CO",
    doi = "10.1103/PhysRevD.105.095029",
    journal = "Phys. Rev. D",
    volume = "105",
    number = "9",
    pages = "095029",
    year = "2022"
}

@article{Caprini:2024hue,
    author = "Caprini, Chiara and Jinno, Ryusuke and Lewicki, Marek and Madge, Eric and Merchand, Marco and Nardini, Germano and Pieroni, Mauro and Roper Pol, Alberto and Vaskonen, Ville",
    collaboration = "LISA Cosmology Working Group",
    title = "{Gravitational waves from first-order phase transitions in LISA: reconstruction pipeline and physics interpretation}",
    eprint = "2403.03723",
    archivePrefix = "arXiv",
    primaryClass = "astro-ph.CO",
    reportNumber = "LISA-COSWG-24-01, CERN-TH-2024-029",
    doi = "10.1088/1475-7516/2024/10/020",
    journal = "JCAP",
    volume = "10",
    pages = "020",
    year = "2024"
}

@article{Linde:1981zj,
    author = "Linde, Andrei D.",
    title = "{Decay of the False Vacuum at Finite Temperature}",
    reportNumber = "LEBEDEV-81-265",
    doi = "10.1016/0550-3213(83)90072-X",
    journal = "Nucl. Phys. B",
    volume = "216",
    pages = "421",
    year = "1983",
    note = "[Erratum: Nucl.Phys.B 223, 544 (1983)]"
}

@article{Peters:2023asu,
    author = "Peters, Fabian Hervas and Schneider, Aurel and Bucko, Jozef and Giri, Sambit K. and Parimbelli, Gabriele",
    title = "{Constraining hot dark matter sub-species with weak lensing and the cosmic microwave background radiation}",
    eprint = "2309.03865",
    archivePrefix = "arXiv",
    primaryClass = "astro-ph.CO",
    reportNumber = "NORDITA 2023-032",
    doi = "10.1051/0004-6361/202449195",
    journal = "Astron. Astrophys.",
    volume = "687",
    pages = "A161",
    year = "2024"
}

@article{Brito:2015oca,
    author = "Brito, Richard and Cardoso, Vitor and Pani, Paolo",
    title = "{Superradiance}: {New Frontiers in Black Hole
Physics}",
    eprint = "1501.06570",
    archivePrefix = "arXiv",
    primaryClass = "gr-qc",
    doi = "10.1007/978-3-319-19000-6",
    journal = "Lect. Notes Phys.",
    volume = "906",
    pages = "pp.1--237",
    year = "2015"
}

@article{Dolan:2007mj,
    author = "Dolan, Sam R.",
    title = "{Instability of the massive Klein-Gordon field on the Kerr spacetime}",
    eprint = "0705.2880",
    archivePrefix = "arXiv",
    primaryClass = "gr-qc",
    doi = "10.1103/PhysRevD.76.084001",
    journal = "Phys. Rev. D",
    volume = "76",
    pages = "084001",
    year = "2007"
}

@article{Baumann:2018vus,
    author = "Baumann, Daniel and Chia, Horng Sheng and Porto, Rafael A.",
    title = "{Probing Ultralight Bosons with Binary Black Holes}",
    eprint = "1804.03208",
    archivePrefix = "arXiv",
    primaryClass = "gr-qc",
    reportNumber = "DESY-18-060, DESY 18-060",
    doi = "10.1103/PhysRevD.99.044001",
    journal = "Phys. Rev. D",
    volume = "99",
    number = "4",
    pages = "044001",
    year = "2019"
}

@article{Tan:2024nbx,
    author = "Tan, Xiu-hui and Xia, Jun-qing",
    title = "{Revisiting bounds on primordial black hole as dark matter with X-ray background}",
    eprint = "2404.17119",
    archivePrefix = "arXiv",
    primaryClass = "astro-ph.CO",
    doi = "10.1088/1475-7516/2024/09/022",
    journal = "JCAP",
    volume = "09",
    pages = "022",
    year = "2024"
}

@article{DelaTorreLuque:2024qms,
    author = "De la Torre Luque, Pedro and Koechler, Jordan and Balaji, Shyam",
    title = "{Refining Galactic primordial black hole evaporation constraints}",
    eprint = "2406.11949",
    archivePrefix = "arXiv",
    primaryClass = "astro-ph.HE",
    reportNumber = "KCL-2024-34",
    doi = "10.1103/PhysRevD.110.123022",
    journal = "Phys. Rev. D",
    volume = "110",
    number = "12",
    pages = "123022",
    year = "2024"
}

@article{Garcia-Bellido:2021zgu,
    author = "Garcia-Bellido, Juan and Murayama, Hitoshi and White, Graham",
    title = "{Exploring the early Universe with Gaia and Theia}",
    eprint = "2104.04778",
    archivePrefix = "arXiv",
    primaryClass = "hep-ph",
    reportNumber = "IFT-UAM/CSIC-2021-038",
    doi = "10.1088/1475-7516/2021/12/023",
    journal = "JCAP",
    volume = "12",
    number = "12",
    pages = "023",
    year = "2021"
}

@article{Janssen:2014dka,
    author = "Janssen, Gemma and others",
    editor = "Bourke, Tyler L. and others",
    title = "{Gravitational wave astronomy with the SKA}",
    eprint = "1501.00127",
    archivePrefix = "arXiv",
    primaryClass = "astro-ph.IM",
    doi = "10.22323/1.215.0037",
    journal = "PoS",
    volume = "AASKA14",
    pages = "037",
    year = "2015"
}

@article{Boudaud:2018hqb,
    author = "Boudaud, Mathieu and Cirelli, Marco",
    title = "{Voyager 1 $e^\pm$ Further Constrain Primordial Black Holes as Dark Matter}",
    eprint = "1807.03075",
    archivePrefix = "arXiv",
    primaryClass = "astro-ph.HE",
    doi = "10.1103/PhysRevLett.122.041104",
    journal = "Phys. Rev. Lett.",
    volume = "122",
    number = "4",
    pages = "041104",
    year = "2019"
}

@article{Laha:2019ssq,
    author = "Laha, Ranjan",
    title = "{Primordial Black Holes as a Dark Matter Candidate Are Severely Constrained by the Galactic Center 511 keV $\gamma$ -Ray Line}",
    eprint = "1906.09994",
    archivePrefix = "arXiv",
    primaryClass = "astro-ph.HE",
    reportNumber = "CERN-TH-2019-099",
    doi = "10.1103/PhysRevLett.123.251101",
    journal = "Phys. Rev. Lett.",
    volume = "123",
    number = "25",
    pages = "251101",
    year = "2019"
}

@article{Cotner:2019ykd,
    author = "Cotner, Eric and Kusenko, Alexander and Sasaki, Misao and Takhistov, Volodymyr",
    title = "{Analytic Description of Primordial Black Hole Formation from Scalar Field Fragmentation}",
    eprint = "1907.10613",
    archivePrefix = "arXiv",
    primaryClass = "astro-ph.CO",
    reportNumber = "IPMU19-0063, YITP-19-31",
    doi = "10.1088/1475-7516/2019/10/077",
    journal = "JCAP",
    volume = "10",
    pages = "077",
    year = "2019"
}

@article{Kawana:2022lba,
    author = "Kawana, Kiyoharu and Lu, Philip and Xie, Ke-Pan",
    title = "{First-order phase transition and fate of false vacuum remnants}",
    eprint = "2206.09923",
    archivePrefix = "arXiv",
    primaryClass = "astro-ph.CO",
    doi = "10.1088/1475-7516/2022/10/030",
    journal = "JCAP",
    volume = "10",
    pages = "030",
    year = "2022"
}

@article{Kawana:2022olo,
    author = "Kawana, Kiyoharu and Kim, TaeHun and Lu, Philip",
    title = "{PBH formation from overdensities in delayed vacuum transitions}",
    eprint = "2212.14037",
    archivePrefix = "arXiv",
    primaryClass = "astro-ph.CO",
    doi = "10.1103/PhysRevD.108.103531",
    journal = "Phys. Rev. D",
    volume = "108",
    number = "10",
    pages = "103531",
    year = "2023"
}

@article{Lewicki:2023ioy,
    author = "Lewicki, Marek and Toczek, Piotr and Vaskonen, Ville",
    title = "{Primordial black holes from strong first-order phase transitions}",
    eprint = "2305.04924",
    archivePrefix = "arXiv",
    primaryClass = "astro-ph.CO",
    doi = "10.1007/JHEP09(2023)092",
    journal = "JHEP",
    volume = "09",
    pages = "092",
    year = "2023"
}

@article{Gouttenoire:2023naa,
    author = "Gouttenoire, Yann and Volansky, Tomer",
    title = "{Primordial black holes from supercooled phase transitions}",
    eprint = "2305.04942",
    archivePrefix = "arXiv",
    primaryClass = "hep-ph",
    doi = "10.1103/PhysRevD.110.043514",
    journal = "Phys. Rev. D",
    volume = "110",
    number = "4",
    pages = "043514",
    year = "2024"
}

@article{Jung:2021mku,
    author = "Jung, Tae Hyun and Okui, Takemichi",
    title = "{Primordial black holes from bubble collisions during a first-order phase transition}",
    eprint = "2110.04271",
    archivePrefix = "arXiv",
    primaryClass = "hep-ph",
    reportNumber = "KEK-TH-2350",
    doi = "10.1103/PhysRevD.110.115014",
    journal = "Phys. Rev. D",
    volume = "110",
    number = "11",
    pages = "115014",
    year = "2024"
}

@article{Murai:2025hse,
    author = "Murai, Kai and Sakurai, Kodai and Takahashi, Fuminobu",
    title = "{Primordial black hole formation via inverted bubble collapse}",
    eprint = "2502.02291",
    archivePrefix = "arXiv",
    primaryClass = "astro-ph.CO",
    reportNumber = "TU-1254",
    doi = "10.1007/JHEP07(2025)065",
    journal = "JHEP",
    volume = "07",
    pages = "065",
    year = "2025"
}

\end{document}